# EVOC: A Computer Model of the Evolution of Culture

Liane Gabora (liane.gabora@ubc.ca)
University of British Columbia
Department of Psychology
Okanagan campus, 3333 University Way
Kelowna BC, V1V 1V7, CANADA

**Abstract**

EVOC is a computer model of the EVOlution of Culture. It consists of neural network based agents that invent ideas for actions, and imitate neighbors' actions. EVOC replicates using a different fitness function the results obtained with an earlier model (MAV), including (1) an increase in mean fitness of actions, and (2) an increase and then decrease in the diversity of actions. Diversity of actions is positively correlated with number of needs, population size and density, and with the erosion of borders between populations. Slowly eroding borders maximize diversity, fostering specialization followed by sharing of fit actions. Square (as opposed to toroidal) worlds also exhibit higher diversity. Introducing a leader that broadcasts its actions throughout the population increases the fitness of actions but reduces diversity; these effects diminish the more leaders there are. Low density populations have less fit ideas but broadcasting diminishes this effect.

## Introduction

In what sense does culture evolve? Is it possible to distil the underlying logic of the process by which ideas adapt and build on one another in the minds of interacting individuals, in the way that Holland's (1975) genetic algorithm, or GA, distilled the underlying logic of natural selection?

'EVOlution of Culture', or EVOC, is an elaboration of Meme and Variations, or MAV (Gabora, 1994, 1995), the earliest computer program to model culture as an evolutionary process in its own right (as opposed to modeling the interplay of cultural and genetic evolution as in (Hutchins & Hazelhurst, 1991)). MAV was inspired by the GA, a search technique that finds solutions to complex problems by generating a 'population' of candidate solutions through processes akin to mutation and recombination, selecting the best, and repeating until a satisfactory solution is found. Although MAV inspired the incorporation of cultural phenomena (such as imitation, knowledge-based operators, and mental simulation) into evolutionary search algorithms (e.g. Krasnogor & Gustafson, 2004), the goal behind MAV was not to solve search problems, but simply to gain insight into how ideas evolve. It used neural network based agents that could (1) invent new ideas by modifying previously learned ones, (2) evaluate ideas, (3) implement ideas as actions, and (4) imitate ideas implemented by neighbors. Agents did not evolve in a biological sense—they neither died nor had offspring—but did in a cultural sense, by generating and sharing ideas for actions. The approach can thus be contrasted with computer models of the interaction between biological evolution and individual learning (Best, 1999, 2006; Higgs, 2000; Hinton & Nowlan, 1987; Hutchins & Hazelhurst, 1991).

MAV successfully modeled how 'descent with modification' could occur in a cultural context, but it had limitations arising from the outdated methods used to program it. Moreover, although the generation of new ideas in MAV capitalized on acquired knowledge, the name 'Meme and Variations' implied acceptance of the idea that novelty is generated randomly, and that culture evolves through a Darwinian process operating on discrete units of culture, or 'memes'. Problems with memetics and other Darwinian approaches to culture have become increasingly apparent (Boone & Smith, 1998; Fracchia & Lewontin, 1999; Gabora, 2004, 2006, 2008; Jeffreys, 2000). One problem is that since natural selection prohibits inheritance of acquired traits, Darwinian approaches must assume that elements of culture are expressed in the same form as that in which they are acquired. In culture, however, 'acquired' change—that is, modification to ideas between the time they are learned and the time they are expressed—is unavoidable. Because ideas cohabit a distributed memory with a multitude of other ideas, their meanings, associations, and implications are constantly revised. EVOC takes a step toward modeling this by allowing agents to have multiple needs that require different actions to be fulfilled.

Other experiments carried out with EVOC but not possible to carry out with MAV investigate how cultural evolution is affected by leadership, and by the affordances of the agents' world, such as (1) world shape and size, (2) population density, and (3) the effect of borders that impede information flow, and potentially erode with time.

## Architecture

EVOC consists of an artificial society of neural network based agents in a two-dimensional grid-cell world. It is written in Joone, an object oriented programming environment, using an open source neural network library written in Java. This section describes the key components of the agents and the world they inhabit.

### The Agent

Agents consist of (1) a neural network, which encodes ideas for actions and detects trends in what constitutes a fit action, and (2) a body, which implements actions. In MAV there was only one need—to attract a mate. Thus actions were limited to gestures that attract mates. In EVOC agents can also engage in tool-making actions.

**The Neural Network.** The core of an agent is an autoassociative neural network, as shown in Figure 1. It is composed of six input nodes that represent concepts of body parts (LEFT ARM, RIGHT ARM, LEFT LEG, RIGHT LEG, HEAD, and HIPS), six matching output nodes, and six hidden nodes that represent more abstract concepts (LEFT, RIGHT, FORELIMB, HINDLIMB, SYMMETRY and MOVEMENT). Input nodes and output nodes are connected to hidden nodes of which they are instances (e.g. RIGHT FORELIMB is connected to RIGHT.) Activation of any input node activates the MOVEMENT hidden node. Same-direction activation of symmetrical input nodes (e.g. positive activation—which represents upward motion—of both forelimbs) activates the SYMMETRY node.

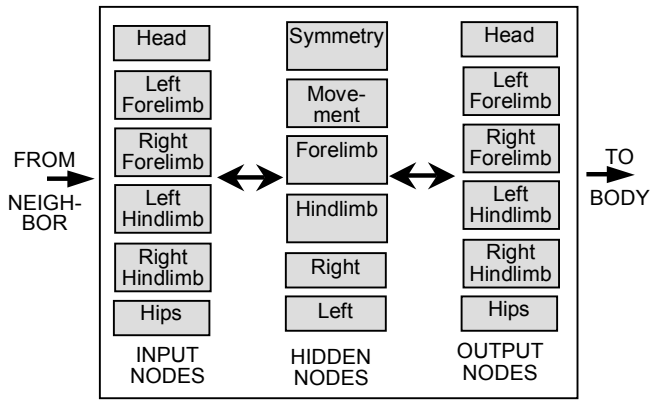

Figure 1. The neural network. See text for details.

The neural network learns ideas for actions. An idea is a pattern consisting of six elements that dictate the placement of the six body parts. Learning and training of the neural network is as per (Gabora, 1995). During imitation, the input is the action implemented by a neighbor. During invention, the pattern of activation on the output nodes is fed back to the input nodes, and change is biased according to the activations of the SYMMETRY and MOVEMENT nodes. In EVOC, the neural network can also be turned off to compare results with a data structure that cannot detect trends, and thus invents ideas merely at random.

**The Body.** If the fitness of an action is evaluated to be higher than that of any action learned thus far, it is copied from the input/output nodes of the neural network that represent *concepts of* body parts to a six digit array that contains representations of *actual* body parts, referred to as the *body*. Since it is useful to know how many agents are doing essentially the same thing, when node activations are translated into limb movement they are thresholded such that there are only three possibilities for each limb: stationary, up, or down. Six limbs with three possible positions each gives a total of 729 possible actions. Only the action that is currently implemented by an agent's body can be observed and imitated by other agents.

**The Fitness Function(s)**

Agents evaluate the effectiveness of their actions according to how well they satisfy needs using a pre-defined equation referred to as a *fitness function*. Agents have two possible needs. The fitness of an action with respect to the need to attract mates is referred to as as $F_1$, and it is calculated as in (Gabora, 1995). $F_1$ rewards actions that make use of trends detected by the symmetry and movement hidden nodes and used by knowledge-based operators to bias the generation of new ideas. $F_1$ generates actions that are relatively realistic mating displays, and exhibits a cultural analog of *epistasis*. In biological epistasis, the fitness conferred by the allele at one gene depends on which allele is present at another gene. In this cognitive context, epistasis is present when the fitness contributed by movement of one limb depends on what other limbs are doing.

The fitness of an action with respect to the second need, the need to make tools, uses a second fitness function, $F_2$, and is calculated as follows. The relevant variables are:

$a_{LH}$ = activation of LEFT HINDLIMB output node
$a_{RH}$ = activation of RIGHT HINDLIMB output node
$a_h$ = activation of HEAD output node
$L$ = 1 if $a_{LH}$ = -0.5, else 0
$R$ = 1 if $a_{RH}$ = -0.5, else 0
$H$ = 1 if $a_h \neq 0$, else $H$ = 0
$c$ = 2.5

$$F_2 = c(L + R + 2H)$$

The constant allows for a maximum fitness of 10 (which is also the maximum fitness using $F_1$). $F_2$ rewards actions in which the head moves (to scan tool), arms either move (to modify tool) or don't (to hold tool), and feet are stationary.

To simulate both needs the fitness functions are combined as follows, where $y$ and $z$ are user-defined variables that allow for differing weightings of the two needs:

$$F_{1+2} = 0.5(yF_1 + zF_2)$$

**The World**

MAV allowed only worlds that were square and toroidal, or 'wrap-around' (such that agents at the left border that attempt to move further left appear on the right border). Moreover, the world was always maximally densely populated, with one agent per cell. In EVOC the world can assume any shape, and be as sparsely or densely populated as required, with agents placed in any configuration. EVOC also allows the creation of complete or semi-permeable permanent or eroding borders that decrease the probability of imitation along a frontier.

**Incorporation of Cultural Phenomena**

Agents incorporate the following phenomena characteristic of cultural evolution as parameters that can be turned off or on (in some cases to varying degrees):

- **Imitation**. Ideas for how to perform actions spread when agents copy neighbors' actions. This enables them to share effective, or 'fit', actions.
- **Invention**. This code enables agents to generate new actions by modifying their initial action or a previously invented or imitated action, as in (Gabora, 1995).
- **Knowledge-based Operators**. Since a new action (or, in invention, new idea for an action) is not learned unless it is fitter than the currently implemented action, new actions provide valuable information about what constitutes an effective idea. This information is used by knowledge-based operators to probabilistically bias invention such that new ideas are generated strategically as opposed to randomly. For example, if successful actions tend to be symmetrical (e.g. left arm moves to the right and right arm moves to the left), the probability increases that new actions are symmetrical. Also, if movement is generally beneficial, the probability increases that new actions involve movement of more body parts. (See (Gabora, 1995) for further details.)
- **Mental simulation**. Before implementing an idea as an action, agents use the fitness function to assess how fit the action would be if it *were* implemented.

**A Typical Run**

Each iteration, every agent has the opportunity to (1) acquire an idea for a new action, either by *imitation*, copying a neighbor, or by *invention*, creating one anew, (2) update the knowledge-based operators, and (3) implement a new action. To invent a new idea, for each node of the idea currently represented on the input/output layer of the neural network, the agent makes a probabilistic decision as to whether change will take place, and if it does, the direction of change is stochastically biased by the knowledge-based operators. If the new idea has a higher fitness than the currently implemented idea, the agent learns and implements the action specified by that idea. To acquire an idea through imitation, an agent randomly chooses one of its neighbors, and evaluates the fitness of the action the neighbor is implementing. If its own action is fitter than that of the neighbor, it chooses another neighbor, until it has either observed all of its immediate neighbors, or found one with a fitter action. If no fitter action is found, the agent does nothing. Otherwise, the neighbor's action is copied to the input layer, learned, and implemented.

Fitness of actions starts out low because initially all agents are immobile. Soon some agent invents an action that has a higher fitness than doing nothing, and this action gets imitated, so fitness increases. Fitness increases further as other ideas get invented, assessed, implemented as actions, and spread through imitation. The diversity of actions initially increases due to the proliferation of new ideas, and then decreases as agents hone in on the fittest actions.

**The Graphic User Interface**

The graphic user interface (GUI) makes use of the open-source charting project, JFreeChart, enabling variables to be user defined at run time, and results to become visible as the computer program runs. The topmost output panel using the mating fitness function ($F_1$) is shown in Figure 2. At the upper left one specifies the *Invention to Imitation Ratio*. This refers to the probability that a given agent, on a given iteration, invents a new idea for an action, versus the probability that it imitates a neighbor's action. Below it is *Rate of Conceptual Change*, where one specifies the degree to which a newly invented idea differs from the one it was based on. Below that is *Number of Agents*, which allows the user to specify the size of the artificial society. Below that is where one specifies *Number of Iterations*, i.e. the duration of a run. The agents that make up the artificial society can be accessed individually by clicking the appropriate cell in the grid on the upper right. This enables one to see such details as the action currently implemented by a particular agent, or the fitness of that action. The graphs at the bottom plot the mean idea fitness and diversity of ideas, in this case using $F_1$ only, i.e. the need to attract a mate. Tabs shown at the top give access to other output panels of the GUI.

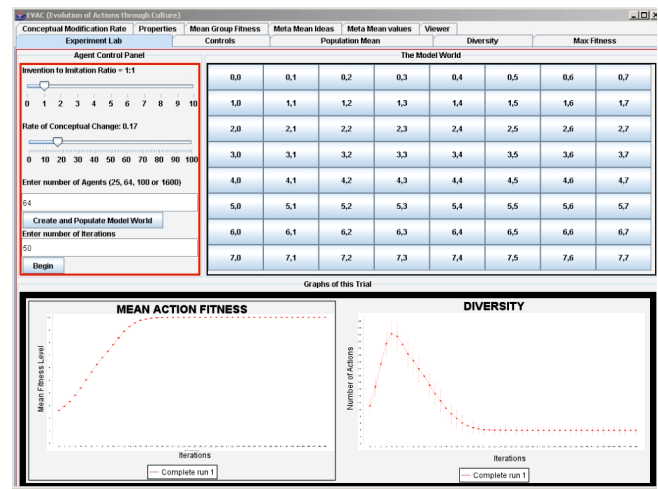

Figure 2. Output panel of GUI using $F_1$. See text for details.

**Replication of Key MAV Results**

EVOC closely replicates the results of experiments conducted with MAV (Gabora, 1995). The graph on the bottom left of Figure 2 shows the increase in fitness of actions. The graph on the bottom right of Figure 2 shows the initial increase and then decrease in the diversity of actions. Other MAV results that are replicated with EVOC include:

- Fitness increases most quickly with an invention to imitation ratio of approximately 2:1.
- For the agent with the fittest actions, however, the *less* it imitates, the better it does.

- Increasing the invention-to-imitation ratio increases the diversity of actions. If increased much beyond 2:1, it takes more than twice as many iterations for all agents to settle on optimal actions.
- As explained earlier, in EVOC, epistasis refers to the situation where the effect on fitness of what one limb is doing depends on what another is doing. As in biology, epistatically linked elements take longer to optimize.
- The program exhibits *drift*—the term biologists use to refer to changes in the relative frequencies of alleles (forms of a gene) as a statistical byproduct of randomly sampling from a finite population (Wright, 1969). With respect to culture, the term pertains not to alleles but to possible forms of a component of an idea (e.g. if the idea is to implement the gesture 'wave', one can do this with one's left hand or one's right).

These results show that concepts from biology are useful in the analysis of cultural change, but that culture also exhibits phenomena that have no biological equivalent.

## Experiments

We now outline the results of new experiments performed with EVOC. Unless stated otherwise, graphs plot the average of 100 runs, and the world consists of 100 cells, one agent per cell, a 1:1 invention to imitation ratio, and a 0.17% probability of change to any body part during invention (since, with six body parts, on average each newly invented action differs from the one it was based on with respect to one body part).

### Effect of Introducing a Different Fitness Function

The first experiment investigated the effect of introducing a different fitness function that fulfills the need to make tools ($F_2$). Figures 3 and 4 show the mean fitness and diversity of actions, respectively, using $F_2$.

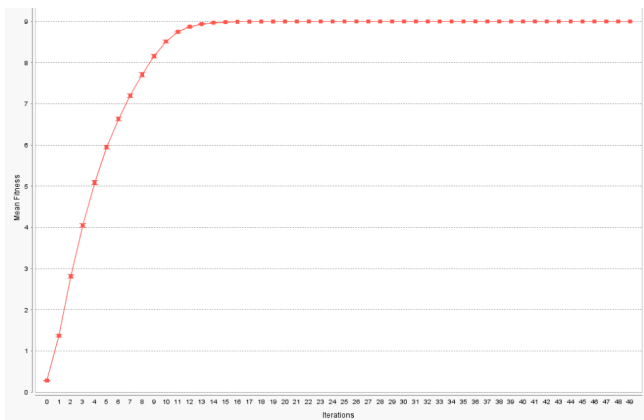

Figure 3. Mean fitness of actions with $F_2$. (Error bars give standard error since we are plotting means of means.)

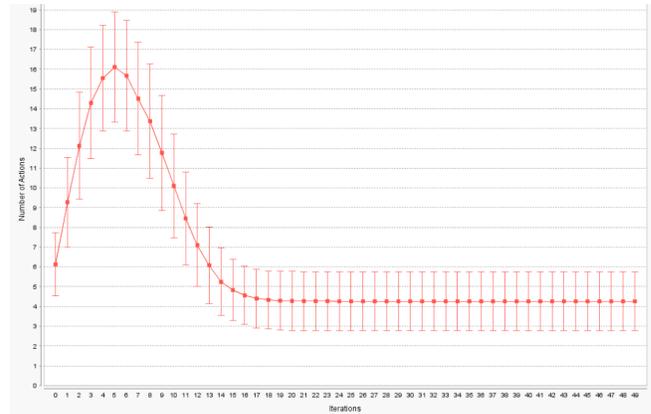

Figure 4. Diversity of actions using $F_2$.

Changing the fitness function did not change the overall pattern of results, as seen by comparing Figures 3 and 4 with the graphs in Figure 2. Mean fitness of actions still increases gradually, and diversity of actions rises and then falls, exhibiting an inverted U-shaped curve, the magnitude of which is a function of population size. However the diversity curve is consistently more lopsided for $F_1$ since it is easy to arrive at a good action but difficult to arrive at an optimal one. This is because optimal actions involve epistasis with $F_1$ but not with $F_2$.

### Multiple Needs

The second experiment investigated the effect of having two needs. The introduction of a second need consistently results in higher diversity, as shown in figure 5.

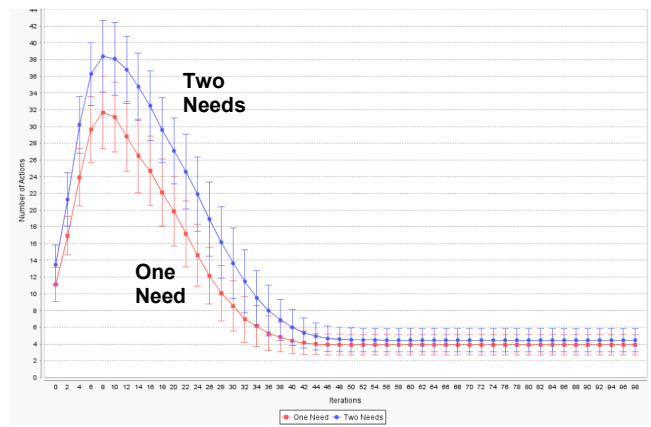

Figure 5. Number of actions with one need versus two.

### Broadcasting

Broadcasting allows the action of a leader to be visible to not just immediate neighbors, but all agents, thereby simulating the effects of media such as public performances, television, radio, or internet, on patterns of cultural change. Each agent adds the broadcaster as a possible source of actions it can imitate. The broadcaster can be specified by the user or chosen at random. Broadcasting can be constant or intermittent. Figure 6 shows the effect on diversity with a randomly chosen broadcaster and constant broadcasting. Broadcasting accelerates convergence on optimal actions but consistently reduces diversity. This effect decreases the more broadcasters there are.

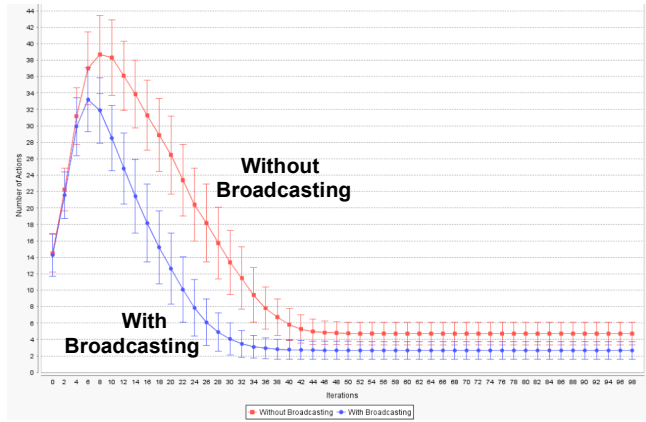

Figure 6. Diversity of actions decreases with broadcasting.

### Effect of Shape of World and Population Density

As in MAV (Gabora, 1995), it is possible to increase both the diversity of actions and the probability of settling on *all* optimally fit actions by increasing either the invention to imitation ratio, or the number of agents. With EVOC this could also be accomplished by changing the shape of the world from toroidal to square. Agents at the edges of a square world have fewer neighbors, and thus more opportunity to retain deviant actions.

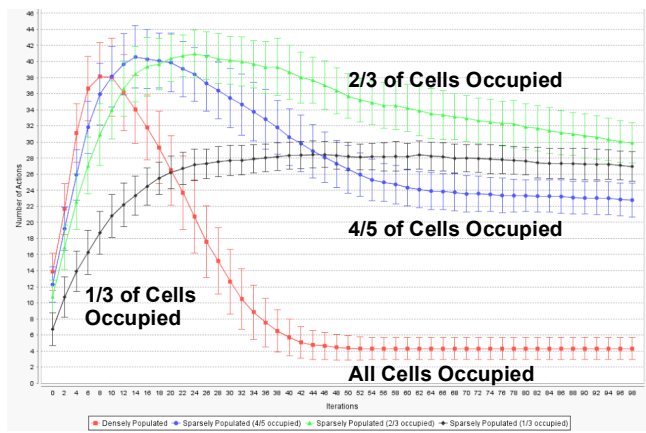

Figure 7. Effect of varying population density on diversity of actions.

Figure 7 compares the diversity of actions over a run with varying population densities. Decreasing the density of agents significantly impairs the ability of the society to converge on only the fittest actions because of the existence of small isolated clusters that are unable to learn from one another and share effective actions. Broadcasting reduces this effect (not shown).

### Semi-permeable Borders

To investigate the impact of impediments to the flow of ideas (e.g. country borders) the effect of reduced probability of imitation between agents on opposite sides of a border was examined. Borders increase latency to converge on fit actions, and increase diversity, by effectively dividing the population. The most interesting results are achieved when borders erode over time such that the probability of imitation by agents on opposite sides is initially zero but increases over the duration of a run, simulating globalization. Eroding barriers foster specialization—honing in on unique solutions—on different sides of the border, followed by sharing of the best to reach a diverse final set. Figure 8 shows the diversity of actions implemented after 4 iterations with an eroding border.

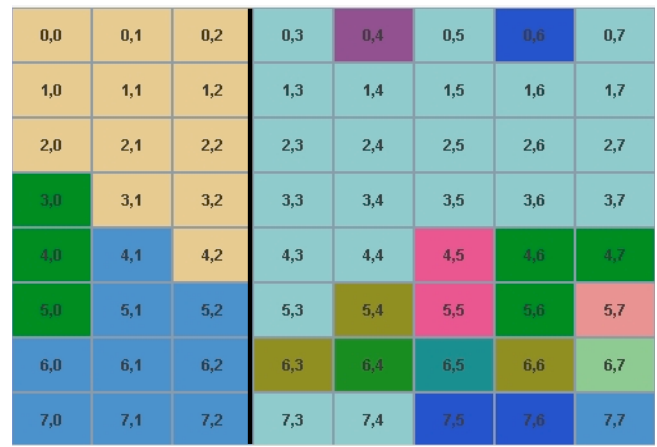

Figure 8. Diversity of actions after four iterations with an eroding barrier between columns 3 and 4. Invention-to-imitation ratio of agents to the right of the border twice is that of agents to the left. Different actions are represented by differently coloured cells. This run used a toroidal, maximally dense 7x7 world.

## Discussion

This paper has given an overview of factors impacting the spread of ideas and behaviors that can be investigated with a computer model of cultural evolution. Results suggest that increasing the number of needs (as happens in a complex society where needs give rise to sub-needs) increases the repertoire of actions, and the benefits of leadership with respect to enhanced fitness of ideas may be tempered by decreased diversity of ideas. This echoes previous simulation findings that leadership can have adverse effects when agents can communicate (Gigliotta, Miglino, & Parisi, 2007). The results also show that properties of the world can have as great an impact on the evolution of culture as properties of the agents themselves.

Further experiments with eroding barriers has potential implications for the impact of free trade on global diversity of ideas, and for investigating the complex relationship between creativity and culture (Kaufman & Sternberg,

2006). Future developments will examine the effect of migration across borders. However, the primary aim of future work will be to examine the distinctively human phenomenon of cultural open-endedness. Although presently agents' actions become more complex and adapted over time, and change is cumulative in that new actions build on existing ones, once agents settle on some subset of optimal actions, the program comes to a standstill. Future versions will use a fitness function that evaluates actions differently depending on the relative strengths of the different needs. The strength of a need will be a function of both how many iterations have passed since execution of an action that satisfied that need, and the degree to which that action satisfied that need. It is expected that the program will not come to a standstill because once an agent has filled one need it will change the kind of action it implements to satisfy another. Moreover to avoid that agents still zero in on predictable subsets of actions that fulfill these needs, future versions of EVOC will incorporate the following:

- ***Context-sensitive concepts***. We plan to move to a more subsymbolic level, incorporating how constellations of activated microfeatures are influenced by context (Aerts & Gabora, 2005a,b; Gabora, Rosch, & Aerts, 2008). This will allow for a richer repertoire of actions.
- ***Chained Actions***. Agents will be allowed to chain actions into arbitrarily long action sequences.
- ***Building Blocks***. Agents will implement actions that cumulatively modify their world using building blocks to create structures that satisfy needs, and add to (or destroy) structures made by others.

With these modifications it is expected that there will no longer be an *a priori* limit to the number or complexity of actions. The role of each of these modifications in bringing about genuine cultural evolution will be assessed. The effort will be judged successful if cultural change is not just cumulative, but cumulative in a way that responds to needs and situations, and open-ended, such that one innovation creates niches for the invention of others (as cars paved the way for the invention of seat belts and gas stations).

## Acknowledgments

Thanks to Martin Denton and Jillian Dicker for work on EVOC. This project is funded by Foundation for the Future and the Social Sciences and Humanities Research Council of Canada (SSHRC).